# Jacquard-woven photonic bandgap fiber displays

I. Sayed[1], J. Berzowska[2], M. Skorobogatiy[1]

[1] Ecole Polytechnique de Montréal, Montréal, Canada, maksim.skorobogatiy@polymtl.ca
[2] XS Labs, Concordia University, Montréal, Canada, joey@xslabs.net
http://www.photonics.phys.polymtl.ca/
http://xslabs.net/karma-chameleon/

## ABSTRACT

We present an overview of photonic textile displays woven on a Jacquard loom, using newly discovered polymer photonic bandgap fibers that have the ability to change color and appearance when illuminated with ambient or transmitted light. The photonic fiber can be thin (smaller than 300 microns in diameter) and highly flexible, which makes it possible to weave in the weft on a computerized Jacquard loom and develop intricate double weave structures together with a secondary weft yarn. We demonstrate how photonic crystal fibers enable a variety of color and structural patterns on the textile, and how dynamic imagery can be created by balancing the ambient and emitted radiation. Finally, a possible application in security ware for low visibility conditions is described as an example.

## 1. Introduction

In recent years, electronic textiles have become a major multi-disciplinary area of research with applications in health, security and safety, communication, and culture industries. Although the field is growing and developing at an accelerated rate, the predominant implementation model still usually consists of layering electronic or mechatronic functionality on top of a textile substrate. Prior work exists in the domain of stitching, weaving, or knitting with conductive yarns to create structures such as electrodes, sensors, or communication lines and subsequently attaching electronic components to that substrate. Few functional yarns (other than conductive or resistive yarns) are currently available commercially to enable functionality such as the display of information, sensing, or energy harnessing in a textile. The ability to integrate the desired functionality on the fundamental level of a fiber remains one of the greatest technological challenges in the development of smart textiles. In this review, we consider photonic fibers as an example of functional yarns that do integrate some functionality at the fiber level.

As indicated by their name, photonic textiles integrate light emitting or light processing elements into mechanically flexible matrix of a woven material, so that appearance or other properties of such textiles can be controlled or sensed with light. Practical implementation of photonic textiles is through integration of specialty optical fibers during the weaving process of textile manufacturing. This approach is quite natural as optical fibers, being long threads of sub-millimeter diameter, are geometrically and mechanically similar to regular textile fibers and ,suitable for similar processing. Various applications of photonic textiles have being researched including large area structural health monitoring and wearable sensing, large area illumination, and garments with unique aesthetic appearance, or flexible and wearable displays.

Optical fibers embedded into woven composites have been applied for in-service structural health monitoring and stress-strain monitoring of industrial textiles and composites [1-3]. Integration of optical fiber-based sensor elements into wearable apparel allows real-time monitoring of bodily and environmental conditions, which is of important to various hazardous civil occupations, including the security and military sector. Examples of such sensor elements can be optical fibers with biologically or chemically activated claddings for bio-chemical detection [4, pp. Bragg gratings and long period gratings [5] for temperature and strain measurements, as well as microbending-based sensing elements for pressure detection [6]. Advantages of optical fiber sensors over other sensor types include resistance to corrosion and fatigue, flexible and lightweight

---

* Corresponding author. Tel.: 1 (514) 340-4711 (3327); Fax: 1 (514) 340-3218
  *E-mail address:* Hmaksim.skorobogatiy@polymtl.ca

nature, immunity to E&M interference, and ease of integration into textiles.

Total Internal Reflection (TIR) fibers modified to emit light sideways [7] have been used to produce emissive fashion items [8, pp. as well as backlighting panels for medical and industrial applications [9,10]. To implement such emissive textiles one typically uses common silica [7] or plastic [11] optical fibers in which light extraction is achieved through corrugation of the fiber surface, or through fiber microbending. Moreover, specialty fibers capable of transverse lasing [12,13] have been demonstrated, with additional applications in security and target identification. Recently, flexible displays based on emissive fiber textiles have received considerable attention due to their potential applications for wearable advertisement [14]. It was noted, however, that such emissive displays are only visible in certain lighting conditions and function as "attention-grabbers" which makes them less suitable for applications that do not require constant user awareness [15]. An alternative to such displays are ambient displays, which are based on non-emissive or weakly emissive elements. An ambient display normally blends in with the environment and is recognized only when the user is aware of it. It is argued that such ambient displays most successfully combine comfort, aesthetics and information streaming capability.

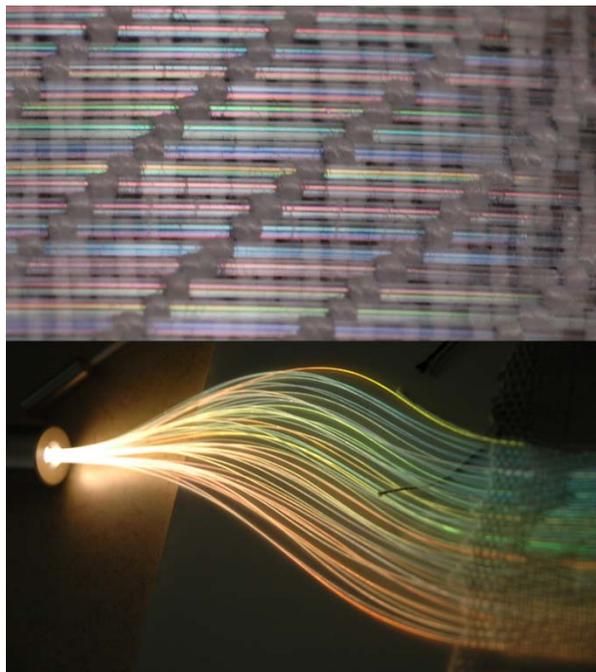

Figure 1. Top: photonic bandgap fibers under ambient illumination show colors due to light interference in their nanostructure. No dyes are used to color the fibers. Bottom: when launching white light into PBG fibers, after few cm from the source, individual fibers become strongly colored by rejecting non-guided colors at the input end.

Recently, a novel type of optical fibers, called photonic crystal fibers (PCFs), has been introduced. In their cross section, such fibers contain either periodically arranged micron-sized air voids [16, pp. or a periodic sequence of micron-sized layers of different materials [17,18]. When illuminated transversally, the spatial and spectral distribution of scattered light from such fibers is quite complex. The fibers appear colored due to optical interference effects in the microstructured region of a fiber. By varying the size and position of the fiber structural elements one can, in principle, design fibers of unlimited unique appearances. Thus, starting with transparent colorless materials, by choosing transverse fiber geometry correctly, one can design the fiber color, translucence and iridescence. This holds several manufacturing advantages, namely that color agents are no longer necessary for the fabrication of colored fibers and the same material combination can be used for the fabrication of fibers with very different designable appearances. Moreover, fiber appearance is very stable over time as it is defined by the fiber geometry rather than by chemical additives such as dyes, which are prone to fading over time. Some photonic crystal fibers guide light using photonic bandgap effect rather than total internal reflection. Intensity of side-emitted light can be controlled by choosing the number of layers in the microstructured region surrounding the optical fiber core. Such fibers always emit a certain color sideways without the need for surface corrugation or microbending, thus promising considerably better fiber mechanical properties compared to TIR fibers adapted for illumination applications. By introducing materials whose refractive index could be changed through external stimuli into the fiber microstructure (for example, liquid crystals at a variable temperature), the spectral position of the fiber bandgap (color of the emitted light) can be varied at will [19]. Finally, as we demonstrate in this work, photonic crystal fibers can be designed to reflect one color when side illuminated and emit another color while transmitting the light. By mixing the two colors one can either tune the color of an individual fiber, or change it dynamically by controlling the intensity of the launched light. This opens new opportunities for the development of photonic textiles with adaptive coloration, as well as wearable fiber-based color displays.

So far, one application of photonic crystal fibers in textiles was demonstrated in the context of distributed detection and emission of mid-infrared radiation (wavelengths of light in a 3-12 µm range) for security applications [17]. the authors used photonic crystal Bragg fibers made of chalcogenide glasses which are transparent in the mid-IR range. Proposed fibers were, however, of limited use for textiles operating in the visible (wavelengths of light in a 0.38-0.75 µm range) due to high absorption of chalcogenide glasses, and a dominant orange-metallic color of the chalcogenide glass. In the visible spectral range, both silica [16] and polymer-based PBG fibers [20] are now available and can be used for textile applications. Currently, however, the cost of textiles based on such fibers would be prohibitively high as the price of such fibers ranges in hundreds of dollars per meter due to the complexity of their fabrication. We believe that the acceptance of photonic crystal fibers by the textile industry will only come if much cheaper fiber fabrication techniques are used. Such techniques can be either extrusion-based [Mignanelli, pp. or should involve simple processing steps requiring limited process control. To this end, our group has developed all-polymer PBG Bragg fibers [18,22] using layer-by-layer polymer deposition and polymer film co-rolling techniques, which are economical and well suitable for industrial scale-up. In this paper we overview the latest advances of our group in fabrication of PBG fiber-based textile panels and review their many intriguing properties [23,24].

## 2. Photonic band gap fiber

In our research we have been fabricating and weaving photonic bandgap (PBG) fibers on a computer-controlled electronic Jacquard loom, in order to produce textile displays with dynamic appearance. In their cross section, PBG Bragg fibers feature periodic sequence of 100's of nano-sized layers of two distinct plastics. Under ambient illumination (Fig. 1, Top) the fibers appear colored due to optical interference in their microstructure. Importantly, no dyes or colorants are used in the fabrication of such fibers, thus making the fibers resistant to color fading. In addition, Bragg fibers guide light in the low refractive index core by photonic bandgap effect (Fig. 1, Bottom), while uniformly emitting a portion of guided color without the need for mechanical perturbations such as surface corrugation or microbending (which is the case in traditional photonic textiles [8,10]), thus making such fibers mechanically superior to standard light emitting fibers. The intensity of side emission is controlled by varying the number of layers in a Bragg reflector. Under white light illumination, the emitted color is very stable over time as it is defined by the fiber geometry rather than by spectral content of the light source. Moreover, Bragg fibers can be designed to reflect one color when side illuminated and to emit another color when transmitting light. By controlling the relative intensities of ambient and guided light, the overall fiber color can be varied, thus enabling passive color changing textiles.

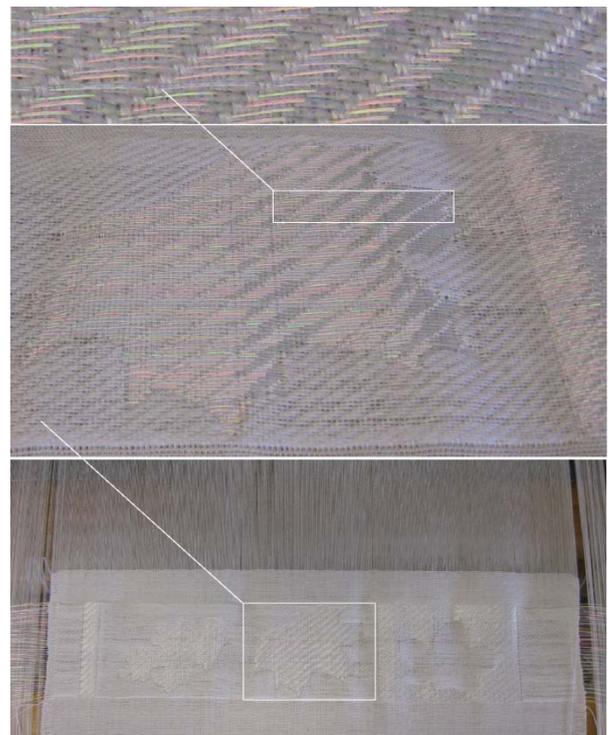

Figure 2. Textile featuring leaf pattern made of cotton thread (warp and weft) and PBG fiber (weft). Leaves are made either with PBG fibers on a cotton background (top left), or with cotton threads on a PBG fiber background (top right).

Fabrication of the solid-core PBG fibers involves layer-by-layer deposition of polymer films, or co-rolling of commercial polymer films around a plastic core rod followed by drawing of these preforms at elevated temperatures [18,23,24]. Material combination of poly methyl methylacrylate (PMMA) and polystyrene (PS) with refractive index contrasts of 1.48/1.6 respectively were used. The core was made of PMMA, while the cladding is a multilayer containing 100's of nano-sized layers (200-1000nm in thickness) of PMMA/PS acting as a color selective all-dielectric mirror. The resultant fiber diameter can be varied between 300 microns and 700 microns. As the fiber color is a result of constructive interference in the reflector layers, it can be adjusted at will by drawing the same preform into fibers of different final diameters.

## 2. Textile reflective properties under ambient illumination

PBG fibers are highly reflective and have the appearance of colored metallic mirrors. A straight PBG fiber exposed to ambient light can "sparkle" when the incidence angle and the angle of observation are properly chosen (see Fig. 2). The sparkling effect can be enhanced by introducing bends into the fiber. The appearance of the reflective textile can be varied dramatically by using different weave patterns, such as the ones enabled by a computerized Jacquard loom. A Jacquard loom allows the weaver to individually address each warp thread so as to create complex weave structures including double weaves. Double weave is a type of woven cloth in which two warps and two sets of weft yarns are interwoven to form a two layered cloth. We use a white cotton warp. We use a white cotton weft and a PBG fiber weft to create individual illuminated designs in the textile display

To systematically study the reflection from PBG fiber textiles, we have prepared a textile sample with a three leaf pattern, each using a variety of weft face double weave structures. We wove the pattern twice, using two different diameters of PBG fibers, 500 and 300 microns. The first sample, shown in Fig. 3 (a), was woven withtthe thicker PBG fibers using a double weave where the warp yarns are separated into two layers so as to bring the PBG fibers to the surface in the leaf pattern and weave them in the back layer in the area around the leaf. In the leaf featured in Fig. 3, the PBG fiber is woven in a 1/7 twill structure in the front layer, while the cotton yarn was woven in a 2/2 pocket weave structure in the back layer. The two layers are connected by reversing the positions of the two layers outside the leaf pattern, but are not connected inside the leaf, which is also known as pocket cloth or pocket weave, since it creates pockets between the two layers.

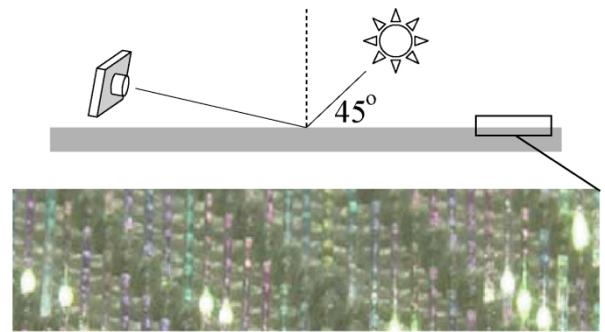

(a)          (b)

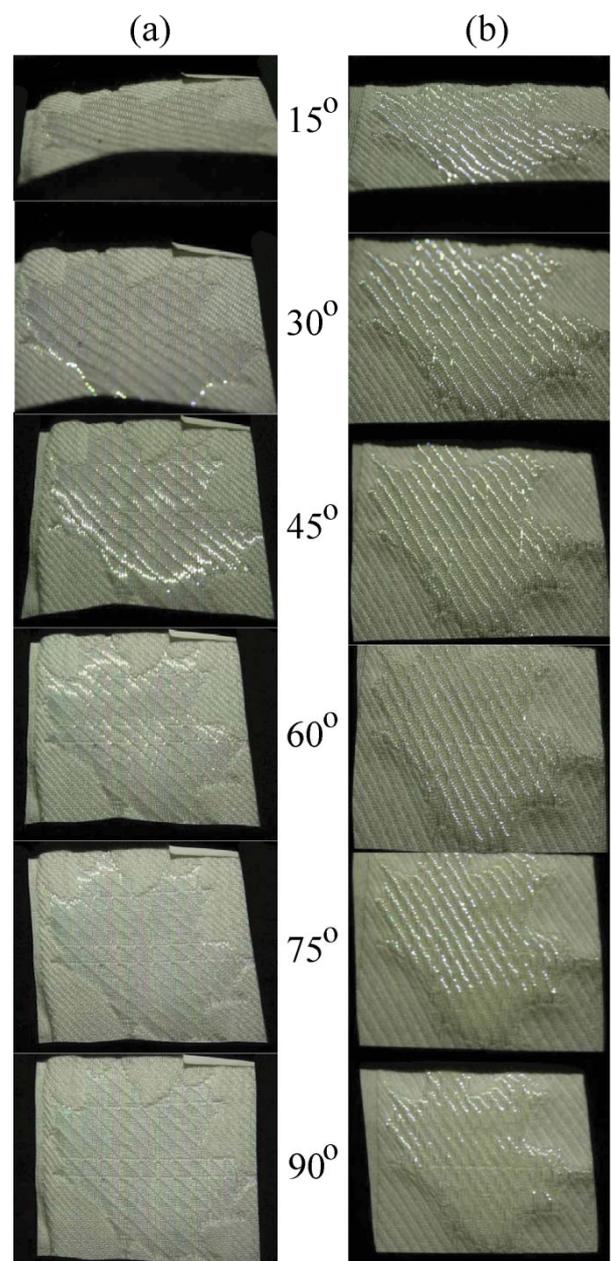

$15^o$

$30^o$

$45^o$

$60^o$

$75^o$

$90^o$

Figure 3. Reflection of ambient light from a PBG fiber-based textile as a function of the angle of observation. The angle of incidence of ambient light is $45^o$. (a) Sample featuring relatively

straight sections of PBG fiber. (b) Sample featuring strongly bent and curled PBG fibers.

The leaf structure features relatively long sections of straight fiber (1-2 cm) placed on the surface of white cotton (see Fig. 2). When reflecting daylight light, textile sections containing Bragg fibers exhibit warm red coloration and have a "sparkly" appearance. We positioned the photo camera 60cm from the center of a sample and took imaged at $15^o$, $30^o$, $45^o$, $60^o$, $75^o$, and $90^o$ inclinations with respect to the textile surface. All the pictures were taken in the dark, with the only illumination source being the Luxeon cool-white LED source (LXK2_PW14_T00 model) placed 60cm from the textile surface at $45^o$ inclination Fig. 3(a). From the photographs we see that the reflection of ambient light from this textile sample is very directional, and that the observed reflection is the strongest when the angle of observation matches the angle of incidence.

Note from Fig. 3 that the color of the reflected LED light appears white, in stark contrast to the color of the reflected daylight (see Fig. 2) which appears reddish. Moreover, when inspecting the same textile sample under magnification in the LED light (see the top part in Fig 3) and in the daylight (see the top part in Fig. 2) one clearly sees that the individual strands are colored in both cases, while the overall hues are different. The main reason for the difference in perceived appearances of the same textile under different illumination condition takes root in the spectral differences of the two illumination sources. Thus, a cool-white spectrum of an LED source features the main maximum in the blue region and a secondary much smaller maximum in the green-red region, while the spectrum of daylight has a much stronger red component.

A second sample, shown in Fig. 3 (b), used the thinner PBG fibers and used similar weave structures as the first. One difference was that since the fibers were thinner, we used a stitched double cloth instead of a pocket cloth. As before, the PBG fiber in the leaf is woven in a 1/7 twill structure in the front layer, while the cotton yarn is woven in a 2/2 pocket weave structure in the back layer. Because the PBG fiber is thinner, it did not provide the same structural rigidity in the pocket structure and we needed to stitch the two layers together by lifting a back end over a face pick in a 1/7 twill pattern, aligned to the 1/7 twill structure in the front layer. This caused increased intersections between the cotton warp yarns and the PBG weft fibers. The tension placed on the weft by the warp threads exerts pressure, which can bend the weft threads. As a result of the relative mechanical properties of the fibers, the thinner PBG fibers were strongly bent at the point of contact with the cotton thread in this weave structure. As seen in Fig. 3(b), this sample scatters light quite differently from the sample shown in Fig. 3(a). Due to the presence of many tight bends in the PBG textile layer, the intensity of the reflected light becomes largely independent of the angle of observation.

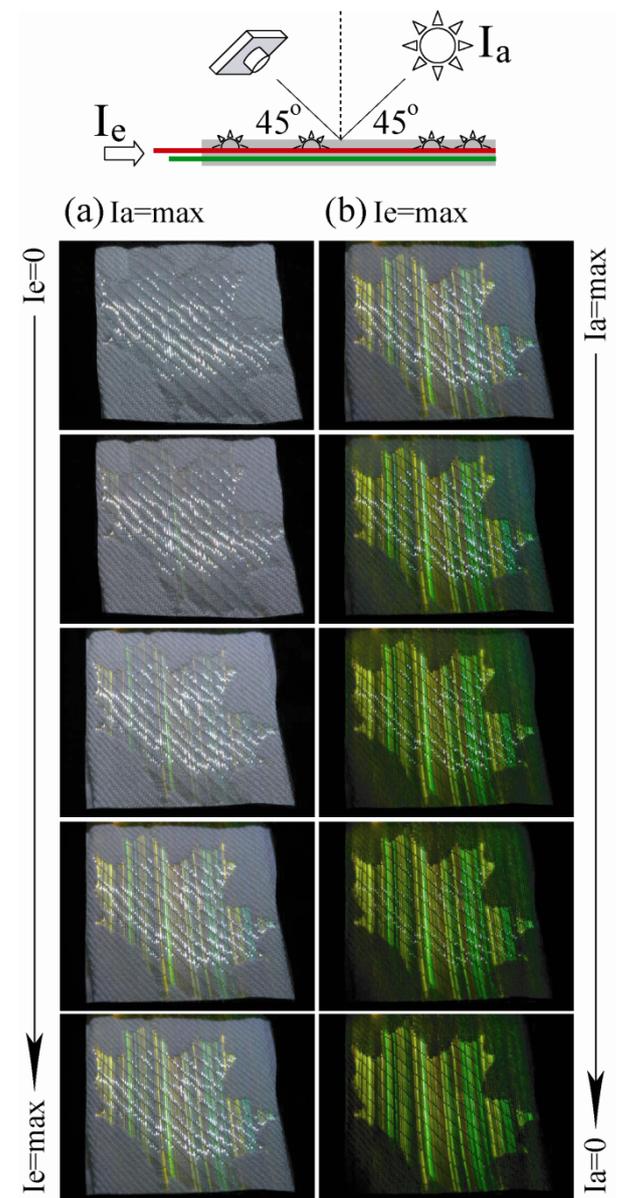

Figure 4. Mixing reflection of the ambient light with emission of the guided light. Left: LED ambient light is at maximum, while intensity of the guided light is varied from zero to its maximal value. Right: intensity of the guided light is at its maximum, while intensity of the LED ambient light is reduced from its maximal value to zero.

## 3. Dynamic animated textile appearance using mixing of ambient and emitted light

To explore variations in textile appearance produced by changes in the relative intensities of ambient illumination and emitted radiation, we have performed the following experiment. The sample under study is a textile panel shown in Fig. 3(a). It was illuminated with a cool-white LED at a 45° angle with intensity Ia, and at the same time, light was launched into the textile with intensity Ie. In Fig. 4(a) we show a sequence of photos which were taken at a fixed intensity of ambient light Ia=const and a variable intensity of light launched into a textile Ie=0→max. We observe that when the intensity of the emitted light is small, the overall appearance of the textile panel is determined by the reflected light. When intensities of the emitted and reflected light are comparable, the textile appearance becomes strongly affected (compare the top and bottom pictures in Fig. 4(a)). Another experiment reflected in a series of photos in Fig. 4(b) was performed by keeping the light launched into a textile at maximal intensity, while gradually reducing the intensity of the ambient light. Note a remarkable dynamics both in color, warmth, iridescence, etc. of a textile provoked by the changes in the illumination conditions.

## 4. Potential applications in worker ware for low visibility conditions

One scenario for PBG textiles considers a civil worker safety garment operating at low light (night) conditions (Fig. 5). In this scenario, the worker is operating at night time, and he/she is wearing an emissive textile panel (see Fig. 5). When no external illumination source is present, one detects a colorful emission from the worker garment. When a car (with headlights turned on) approaches from a distance, the overall appearance of the textile starts changing. The closer the car, the stronger the components of light reflected from the garment. As the overall appearance of the textile is strongly sensitive to a balance between the intensities of emitted and reflected light (see previous section), we expect strong changes in textile appearance under this scenario. In Fig. 5, we demonstrate the results of an experiment in which a camera and a lit camera-mounted LED were placed at various distances from a lit photonic textile panel. All the pictures were taken with the same shutter speed and aperture.

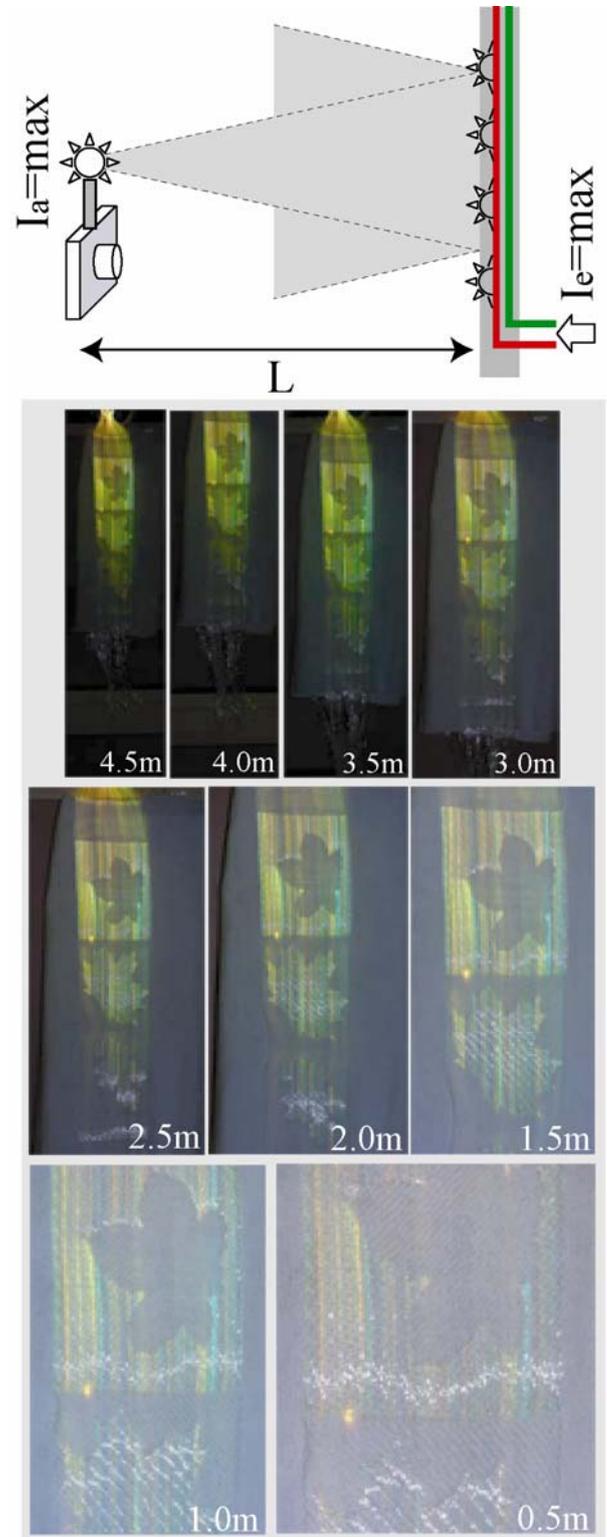

Figure 5. Changes in the textile appearance under a night time scenario of an approaching car with turned on headlights. The emissive textile panel

represents a worker operating at night. Workers appearance changes considerably when a bright illumination source (car headlights) approaches.

When the LED and the camera are placed far from the textile panel (L>4.0 m), the camera could only register the light emitted from the photonic textile. When moving closer towards a textile panel, reflection of the LED light becomes more pronounced. At a distance L~1.5 m, intensities of the reflected and emitted light become comparable, while at smaller distances textile appearance is dominated by reflections. As the human eye is most sensitive to detection of changes in the appearance of objects, a PBG fiber-based photonic textile panel will be highly noticeable in the night time environment.

## 5. Conclusion

We have presented examples of the novel photonic bandgap fiber-based textiles for potential applications in smart cloths, signage and art. It was established that under ambient illumination such textiles appear colored due to optical interference in the microstructure of constitutive optical fibers, with no color dyes needed. Moreover, when launching white light into PBG fibers, all but a single color are rejected, while the remaining color is guided and gradually emitted along the textile length. Note that PBG fibers naturally emit sideways a portion of guided color without the need of any mechanical perturbation, which differentiates them favorably from the competing standard optical fibers. We have also demonstrated that PBG fibers can reflect one color when side illuminated, and emit another color while transmitting light. We then demonstrated that by controlling the relative intensities of the ambient and guided light the overall textile appearance (color, warmth, iridescence, etc.) can be strongly varied. Finally, we have explored the possibility of using PBG fiber-based textiles in the safety worker ware for operation at night time. We have demonstrated that when the external light source, such as car headlights, approaches an emissive PBG fiber-based textile panel, textile appearance can change dramatically, thus being highly noticeable.

**Acknowledgements**

We would like to thank Hexagram research center for giving us the access to Jacquard loom, Ronald Borshan for helping photograph the textiles, Francis Boismenu for his continuing technical support, and Damir Cheremisov and Helene Gingras for weaving assistance. We would like to acknowledge NSERC Media Arts Initiative and Canada Art Council for their financial support of this project.

**REFERENCES** Au, L.T. & Wang, A. 1997, 'The importance of denim wear for university students', *Research Journal of Textile and Apparel*, vol. I, no. 2, pp. 105-113.

[1] Uskokovic, P.S., Miljkovic, M., Krivokuca, M., Putic, S.S., Aleksici, R, 1999, "An intensity based optical fibre sensor for flexural damage detection in woven composits," *Adv. Composits Lett*, vol. 8, pp. 55-58.

[2] D'Amato, E. 2002, "Stress-strain monitoring in textile composites by means of optical fibers," Exp. Techniques & Design in Composite Mat., vol. 221, pp. 245-253 (2002).

[3] Kuang, K.S.C., Cantwell, W.J., 2003, "Detection of impact damage in thermoplastic-based glass fiber composites using embedded optical fiber sensors," J. Thermoplastic Composite Mat., vol. 16, pp. 213-229.

[4] El-Sherif, M.A., Yuan, J.M., MacDiarmid, A. 2000, "Fiber optic sensors and smart fabrics," J. Intelligent Mat. Systems & Struct., vol. 11, pp. 407-414.

[5] S. Ghosh, C. Amidei, K. Furrow, "Development of a sensor-embedded flexible textile structure for apparel or large area applications," Indian J. Fibre \& Textile Res., vol. 30, pp. 42-48 (2005).

[6] Zheng, Y.H., Pitsianis, N.P., Brady, D.J., 2006, "Nonadaptive group testing based fiber sensor deployment for multiperson tracking," IEEE Sens. J. Color Res. Appl., vol. 6, pp. 490-494.

[7] Spigulis, J., Pfafrods, D., Stafekis, M., Jelinska-Platace, W., 1997, "The 'glowing' optical fibre designs and parameters," Proc. SPIE, vol. 2967, pp. 231-6.

[8] www.lumigram.com

[9] Selem, B., Rothmaier, M., Camenzind, M., Khan, T., Walt, H., 2007, "Novel flexible light diffuser and irradiation properties for photodynamic therapy," J. Biomed. Opt., vol. 12, pp. 034024.

[10] www.lumitex.com

[11] Harlin, A., Makinen, M., Vuorivirta, A., 2003, "Development of polymeric optical fibre fabrics as illumination elements and textile displays," AUTEX Res. J., vol. 3.
[12] Balachandran, R.M., Pacheco, D.P., Lawandy, N.M., 1996, "Photonic textile fibers," Applied Optics, vol. 35, pp. 91-94.
[13] Shapira, O., Kuriki, K., Orf, N.D., Abouraddy, A.F., Benoit, G., Viens, J.F., Rodriguez, A., Ibanescu, M., Joannopoulos, J.D., Fink, Y., 2006, "Surface-emitting fiber lasers," Opt. Express, vol. 14, pp. 3929-3935.
[14] Koncar, V, 2005, "Optical fiber fabric displays," Opt. Photon. News, vol. 16, p. 40.
[15] Wakita, A., Shibutani, M., 2006, "Mosaic Textile: Wearable ambient display with non-emissive color-changing modules," Proc. ACE 06.
[16] Knight, J.C., Broeng, J., Birks T.A., Russell, P.J., 1998, "Photonic band gap guidance in optical fibers," Science 282, 1476-1478.
[17] Hart, S.D., Maskaly, G.R., Temelkuran, B., Prideaux, P.H., Joannopoulos, J.D., Fink, Y., 2002, "External reflection from omnidirectional dielectric mirror fibers," Science, vol. 296, pp. 510-513.
[18] Dupuis, A., Guo, N., Gauvreau, B., Hassani, A., Pone, E., Boismenu, F., Skorobogatiy, M., 2007, "Guiding in the visible with "colorful" solid-core Bragg fibers," Opt. Lett., vol. 32, pp. 2882-2884.
[19] Larsen, T.T., Bjarklev, A., Hermann, D.S., Broeng, J., 2003, "Optical devices based on liquid crystal photonic bandgap fibres," Opt. Express, vol. 11, pp. 2589-2596.
[20] Argyros, A., Bassett, I., Eijkelenborg, M., Large, M., Zagari, J., Nicorovici, N.A., McPhedran, R., de Sterke, C.M., 2001, "Ring structures in microstructured polymer optical fibres," Opt. Express, vol. 9, pp. 813-820.
[21] Mignanelli, M., Wani, K., Ballato, J., Foulger, S., Brown, P., 2007, "Polymer microstructured fibers by one-step extrusion," Opt. Express, vol. 15, pp. 6183-6189.
[22] Gao, Y., Guo, N., Gauvreau, B., Rajabian, M., Skorobogata, O., Pone, E., Zabeida, O., Martinu, L., Dubois, C., Skorobogatiy, M., 2006, "Consecutive solvent evaporation and co-rolling techniques for polymer multilayer hollow fiber preform fabrication," J. Mater. Res., vol. 21, pp. 2246-2254.
[23] Gauvreau, B., Guo, N., Schicker, K., Stoeffler, K., Boismenu, F., Ajji, A., Wingfield, R., Dubois, C., Skorobogatiy, M. 2008, 'Color-changing and color tunable photonic bandgap fiber textiles', *Optics Express*, vol. 16, pp. 15677.
[24] Gauvreau, B., Schicker, K., Guo, N., Dubois, C., Wingfield, R., Skorobogatiy, M., 2008 'Color-on-demand photonic textiles', *The Textile Journal*, vol. 125, pp. 70-81